\documentclass{cys}

\usepackage[utf8]{inputenc}
\usepackage[bookmarks=false]{hyperref}
\usepackage{tabularx}
\usepackage[english]{babel}
\usepackage{float}
\usepackage{placeins}

\usepackage{graphicx}
\usepackage{epstopdf} 

\addto\captionsenglish{%
}

\addto\captionsspanish{%
}

\title{Deep Neural Network for Musical Instrument Recognition using MFCCs}

\author{Saranga Kingkor Mahanta$^1$, Abdullah Faiz Ur Rahman Khilji$^2$, Partha Pakray$^2$}

\affil{ 
$^1$ Department of Electronics and Communication Engineering, National Institute of Technology, Silchar, Assam,  \authorcr   
India
\authorcr \authorcr
$^2$ Department of Computer Science and Engineering, National Institute of Technology, Silchar, Assam,  \authorcr   
India
\authorcr  \authorcr
\{saranga\_ug@ece, abdullah\_ug@cse, partha@cse\}.nits.ac.in
\authorcr  \authorcr
}

\begin{document}

\maketitle

\renewcommand{\tablename}{Table}

\begin{abstract}
The task of efficient automatic music classification is of vital importance
and forms the basis for various advanced applications of AI in the musical
domain. Musical instrument recognition is the task of instrument identification by
virtue of its audio. This audio, also termed as the sound vibrations are leveraged by
the model to match with the instrument classes. In this paper, we use an artificial
neural network (ANN) model that was trained to perform classification on twenty
different classes of musical instruments. Here we use use only the mel-frequency
cepstral coefficients (MFCCs) of the audio data. Our proposed model trains on
the full London philharmonic orchestra dataset which contains twenty classes
of instruments belonging to the four families viz. woodwinds, brass, percussion,
and strings. Based on experimental results our model achieves state-of-the-art
accuracy on the same.
\end{abstract}

\begin{keywords} 
Musical Instrument Recognition, Artificial Neural Network, Deep Learning, Multi-class Classification
\end{keywords}

\section{Introduction}
\label{sec:introduction}
Music is an essential part of our daily lives for a majority of the world population. Categorization of music can be based on various parameters like genre, performer or composer. Machine Learning techniques provide numerous ways to perform music categorization as per need. Automatic musical instrument recognition and classification is a non-trivial and practically valuable task as it effectively classifies music with respect to the instrument being played in a faster and cheaper way than carrying it out manually. Even seasoned musicians face difficulty to distinguish between two instruments belonging to the same family, thus manual classification is also prone to errors. Automatic recognition of musical instruments forms the basis of more complex tasks such as melody extraction, music information retrieval, recognizing the dominant instruments from polyphonic audio \cite{essid2005instrument}, and so on. Hence the classification task is vital for subsequent downstream tasks. 

This paper proposes a deep artificial neural network model that efficiently distinguishes and recognizes 20 different classes of musical instruments, even across instruments belonging to the same family. Additionally, an attempt to examine the distinctive potency of MFCCs for the classification task is made. Currently, MFCCs are being widely used as deciding feature for numerous cognition tasks since they take into account the perception of sound with regards to the functioning of the human hearing system, by converting the conventional frequency to the Mel Scale. The results achieved with the simple proposed model on a highly imbalanced dataset without needing to implement data augmentation techniques or explicitly handling the class imbalance bolsters the significance of MFCCs. The promising results can be claimed as a new state-of-the-art on the particular dataset because previous works have considered different subsets of the dataset, while this work considers the entire dataset.

The organization of the paper is as follows. Section \ref{sec:related_works} contains the musical instrument recognition related works and literature survey, Section \ref{sec:dataset} describes the dataset; the pre-processing steps have been explained in Section \ref{sec:pre-pro}; Section \ref{sec:system} is the System Description; the setup and description of our model is explained in Section \ref{sec:setup}; the results and discussions are present under Section \ref{sec:results}; finally, Section \ref{sec:conclusion} concludes the paper with future scopes and potential improvements.

\section{Related Work}
\label{sec:related_works}
Ample research projects have been undertaken to recognize and classify one musical instrument from the other. Numerous feature extraction techniques \cite{deng2008study} have been formulated followed by various machine learning implementations to accurately perform this particular task of classification. In 1999, a study classified 8 different instruments with a 30\% error rate using gaussian mixture models and support vector machines (SVM) \cite{marques1999study}. Hidden markov models have been used by \cite{eichner2006instrument} to classify audio streams among four instruments on a dataset containing 600 recordings. Profound use of The K-Nearest Neighbour classifier to compare performances with other models has been observed as well. The Gaussian and K-NN classifiers were implemented by \cite{eronen2000musical} to classify instruments with a set of 43 feature inputs. Another study implemented the K-NN and SVM on cepstral features of the audio data \cite{gulhane2018identification}. 

With the advent of advancing deep learning methods, recent studies have exploited the power of neural networks to attain excellent levels of classification accuracies. Moreover, deep learning methods exempt researchers from the tedious task of manually extracting multiple features to be fed into a learning model. Convolutional neural networks (CNNs) have been used by \cite{haidar2019music}, \cite{solanki2019music}, and \cite{singh2019implementing} on mel-spectrograms, visual representation of sound signals that encompasses both time and frequency domain features.

A few studies have worked on the London philharmonic orchestra dataset and achieved notable results. ANNs were trained by \cite{toghiani2017musical} on 8 out of the 20 instrument classes currently present in the dataset. They also conducted several comparative experiments on different characteristics of the musical instruments that might have impacted the resulting accuracy. They achieved a maximum accuracy of 93.5\% on the base experiment. A stable precision and recall of 94\% were achieved by \cite{siebertstudy} on the classification performed on 18 classes of the same dataset. Another study achieved a 99\% accuracy using a CNN, however, the model was trained on only 6 classes \cite{siebertstudy}.

With increase in the number of classes, the complexity of the classification increases. Unlike the previously mentioned works that work on a subset of the dataset, our model is trained on all 20 classes of the London philharmonic dataset that are currently available, and achieves a commendable accuracy of 97\%.

\section{Dataset}
\label{sec:dataset}

The proposed model has been trained on all classes of the London philharmonic orchestra dataset\footnote{London philharmonic orchestra all instruments dataset publicly available at \url{https://philharmonia.co.uk/resources/sound-samples/}}. At the time of downloading the dataset, there were a total of 13679 examples divided non-uniformly among 20 classes of musical instruments. 

For each instrument class, except `percussion', the samples are musical notes ranging from A1 (note A of the first octave) to G7 (note G of the seventh octave). The tones have been played with varying dynamics including forte, fortissimo, piano, pianissimo; playing techniques, trills and sustains- staccato, staccatissimo, legato, vibrato, tremolo, pizzicato, ponticello, thus bestowing us with diverse samples for a learning system to generalize over.

\begin{figure*}[ht]
	\centering
    \includegraphics[width=\textwidth, height= 9 cm]{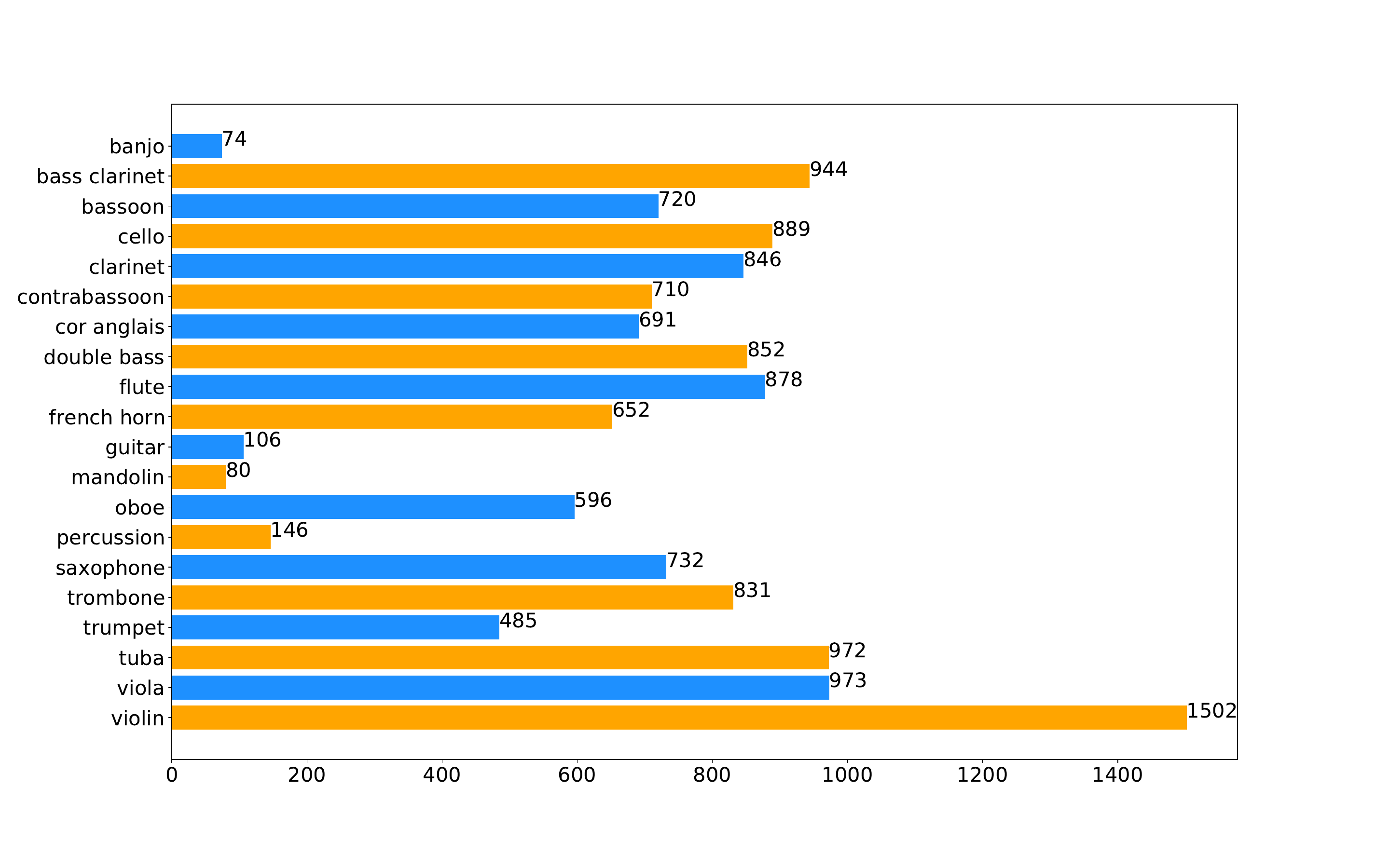}
    \caption{Data distribution among the 20 classes}
    \label{fig:distribution}
\end{figure*}

From Figure \ref{fig:distribution} it can be observed that the dataset is highly imbalanced with `violin' having 1502 examples while `banjo' having only 74.

Additionally, The class named `percussion' has 39 sub-classes of percussion instruments ranging from `agogo bells' to `woodblock'. The sub-classes have too few examples to be treated as individual classes, hence they are consolidated into a single class i.e. `percussion'.

\section{Pre-processing}
\label{sec:pre-pro}

The data was already noise-free and consisted of single instrument tones per example corresponding to the respective class, thus relieving us from performing complex processing procedures. The various steps of pre-processing that were performed have been described in detail in the following sections.

\begin{figure*}[ht]
	\centering
    \includegraphics[width=\textwidth, height=4cm]{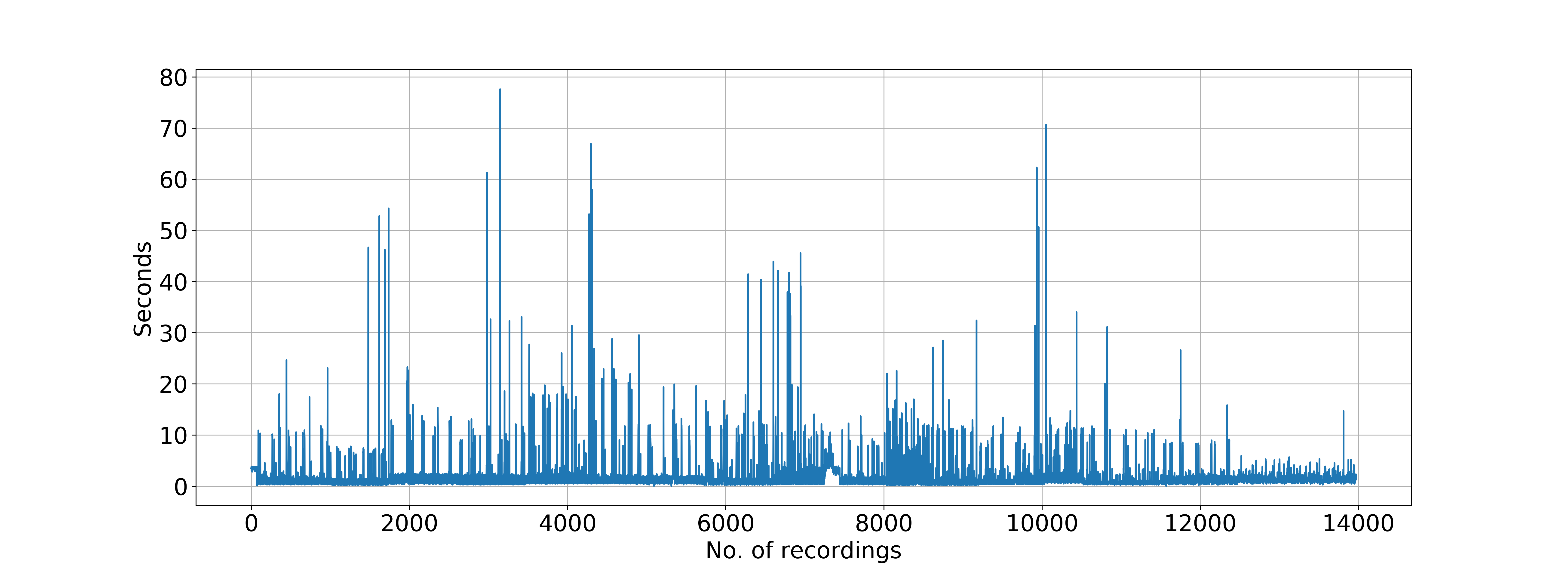}
    \caption{Durations of all examples}
    \label{fig:durations}
\end{figure*}

\subsection{Normalization of Audio Files}
The audio data have durations ranging from 0.078 seconds to 77.06 seconds, as shown in the Figure \ref{fig:durations}. In the proposed model, we use MFCCs of the training examples as input features into the ANN. It is mandatory that the input dimensions must be a constant as the input layer of the ANN has a static number of neurons, thus implying that the MFCC matrices of all the examples need to have a fixed dimensional size. To achieve this, each of the examples must compulsorily have a constant duration resulting in a fixed number of samples when sampled with a constant sampling rate.

\begin{figure}[ht]
	\centering
    \includegraphics[width=\linewidth, height=3.5cm]{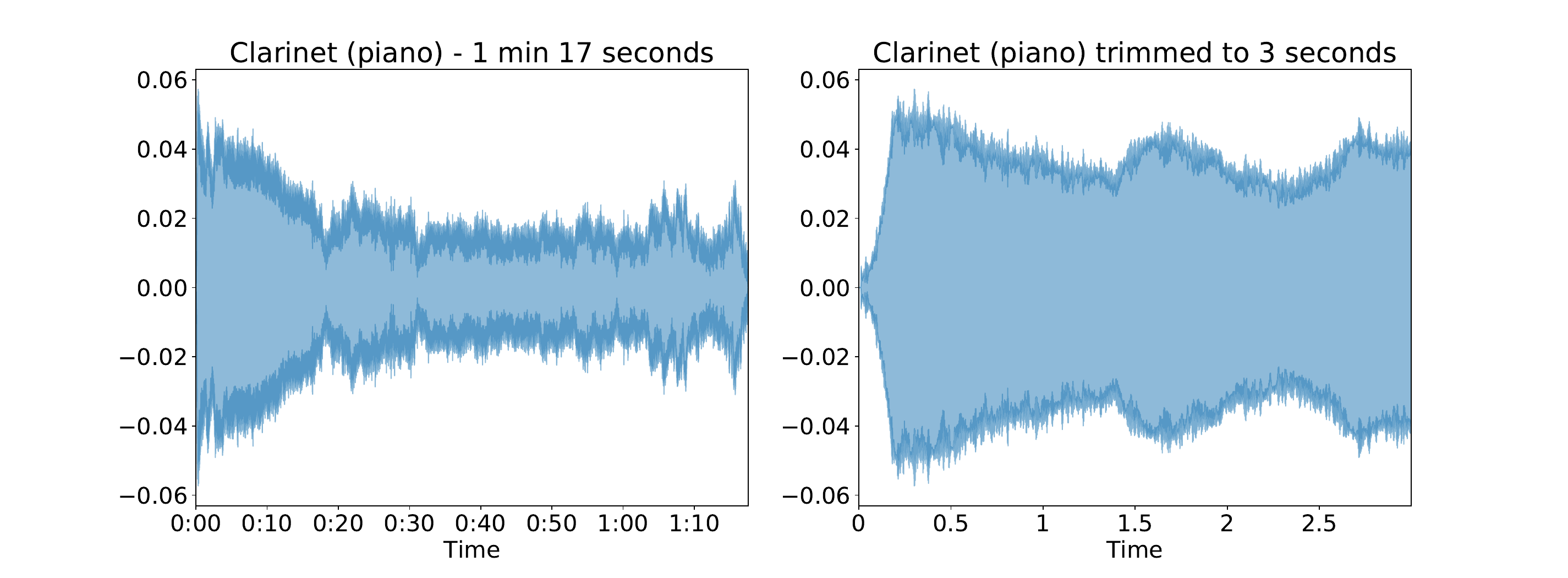}
    \caption{Longest audio clip trimmed to 3 seconds}
    \label{fig:longest}
\end{figure}

\begin{figure}[ht]
	\centering
    \includegraphics[width=\linewidth, height=3.5cm]{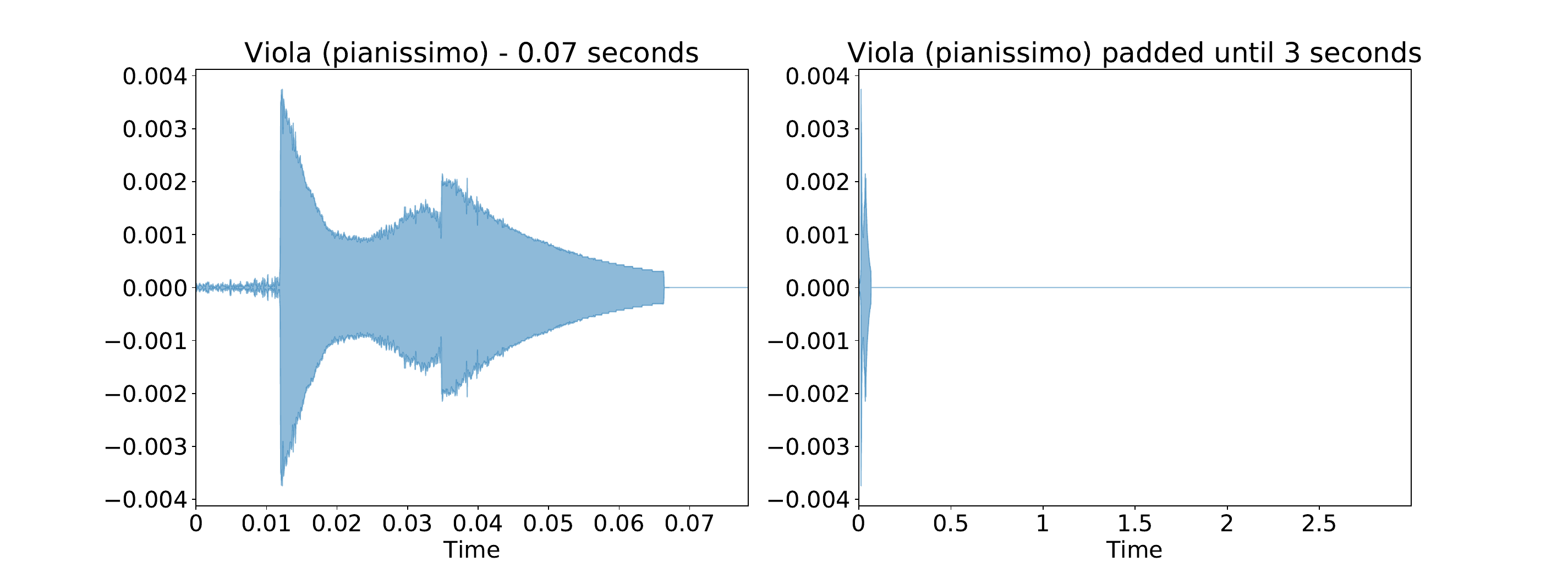}
    \caption{Shortest audio clip padded until 3 seconds}
    \label{fig:shortest}
\end{figure}

Out of the 13679 examples, it was observed that 6342, 11097, 12196, 12550, 12701 examples had durations less than or equal to 1, 2, 3, 4 and 5 seconds respectively. Since a majority of the audio files had a duration lesser than or equal to 3 seconds, a fixed duration of 3 seconds was chosen for all the examples\footnote{3 seconds correspond to 66150 samples for the default sampling rate of 22050 Hz}. As a result, while loading the examples those having a duration of more (Figure \ref{fig:longest}) or less (Figure \ref{fig:shortest}) than 3 seconds were trimmed or padded respectively and finally in the preprocessed data all 13679 examples had a duration from 0 to 3 seconds.

It can be concluded that no promising information was lost from the trimmed samples because all of the audio files' onsets and attacks of the sound take place within the first 3 seconds following which the sound either decays or sustain, i.e the defining features of the audio clip mostly occur before 3 seconds into the signal and most of the examples are almost periodic with minuscule periods. Another reason for not choosing a fixed length of more than 3 seconds is to limit the number of sparse values that result from padding and to reduce dimensional size.

\subsection{Extracting Mel-Frequency Cepstral Coefficients}

MFCCs are useful in identifying the formants and timbre of sound \cite{chakraborty2018improved}, as described in Section \ref{sub-sec:MFCC}. 13 MFCC coefficients were extracted from each frame of each audio file. A single frame contained 2048 samples. A hop length of 512 frames was used for the framing window. The 3 second sound samples resulted in MFCC matrices having dimensions of 13x130. These matrices were then simply flattened and fed into our proposed ANN model.

\section{System Description}
\label{sec:system}

Our approach consists of two principal steps. Extraction of the MFCC features, as shown in Figure \ref{fig:MFCC} from the constant length examples and feeding them into an ANN model to make predictions.

\subsection{Mel-Frequency Cepstral Coefficients}
\label{sub-sec:MFCC}

\begin{figure*}[ht!]
	\centering
    \includegraphics[width=\linewidth]{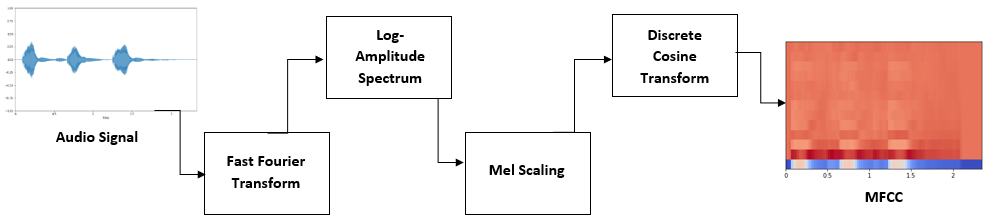}
    \caption{Steps to extract Cepstral Coefficients from an audio signal}
    \label{fig:MFCC}
\end{figure*}

For our approach, we are most interested for our model to be able to distinguish the timbres of tones belonging to different instruments efficiently. Timbre, or tone colour, is a characteristic of sound that can distinguish between two sounds possessing the same intensity, frequency, and duration. Since many examples belonging to different classes play the same note with similar intensity, the timbre becomes a crucial factor to differentiate amongst them. 

The information of the rate of change in spectral bands of a signal is given by its cepstrum. It conveys the different values that construct the formants and timbre of a sound. The cepstral coefficients can be extracted using Equation \ref{cepstrum}.

\begin{equation}
    \label{cepstrum}
    C(x(t))=F^{- 1}(log(F[x(t)]))
\end{equation}

Peaks are observed at periodic elements of the original signal while computing the log of the magnitude of the Fourier transform of the audio signal followed by taking its spectrum by a cosine transformation. The resulting spectrum lies in the quefrency domain \cite{oppenheim2004frequency}. Humans perceive amplitude logarithmically, hence conversion to the Log-Amplitude Spectrum is perceptually relevant, as shown in Figure \ref{fig:MFCC}. Mel scaling is performed on it by using Equation \ref{mel} on frequencies measured in Hz.

\begin{equation}\label{mel}
Mel(f)=2595 * log(1+f/700)
\end{equation}

Human beings can distinguish small changes in speech at lower frequencies. The Mel Scale captures these tiny differences to draw relations with the hearing process of humans. The discrete cosine transformation is mostly applied instead of an inverse fourier transform as shown by Equation \ref{cepstrum}. Since the former provides real-valued coefficients and also decorrelates energy in different mel bands.

For our model, the traditional 13 MFCC coefficients were chosen per frame of each example since the lower end of the quefrency axis of the cepstrum contains the most relevant information to our particular task viz. formants, spectral envelope and timbre. Towards the higher end of the same, information related to the glottal pulse can be obtained, which is not very defining for the same.

\subsection{The Artificial Neural Network}
A deep ANN comprises numerous layers containing a varying number of neurons terminating into the output layer having an equal number of neurons as the number of classes with regards to a classification task; in our case 20. The feature values are multiplied with weights and added with bias terms. These weights are then updated after each epoch consisting of a forward and backward propagation with respect to a chosen loss function, as input feature values traverse through the neurons of subsequent layers. The neurons also apply an activation function on the computed values to introduce non-linearity and selectivity in the network. These activation functions distinguish a neural network from a regular linear regression model. 

In our model, we use the Rectified Linear Unit (ReLU) activation function for all the hidden layers. It simply activates the neurons containing a positive value after the aforementioned computations. It is given by Equation \ref{relu}.

\begin{equation}\label{relu}
y = max(0, x)
\end{equation}

The current problem being a multi-class classification, the Softmax function is used in the output layer. It provides the confidence scores of each class using the Equation \ref{softmax}. The scores add up to 1. The class having the highest confidence score is the model's predicted class for a particular set of input features.

\begin{equation}
    \sigma(z)_i = \frac{e^{z_i}}{\sum^{K}_{j=1}e^{z_j}}
    \label{softmax}
\end{equation}
\section{Experimental Setup}
\label{sec:setup}

\begin{figure*}[ht!]
	\centering
    \includegraphics[width=\textwidth,height=7cm]{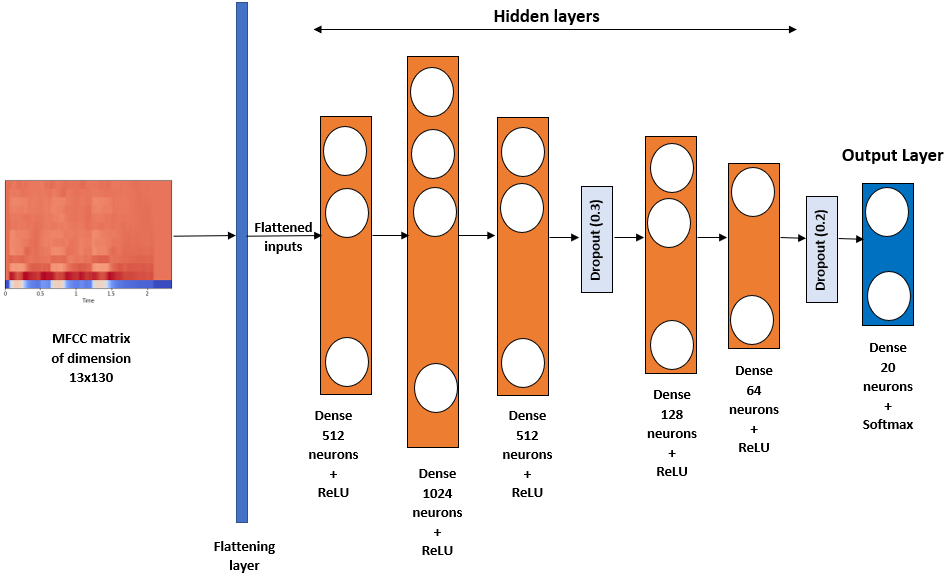}
    \caption{Proposed model architecture}
    \label{fig:ANN}
\end{figure*}

Our approach uses an ANN whose architecture is described in Figure \ref{fig:ANN}. The initial flattening layer flattens a 13x130 MFCC matrix into a single-dimensional layer having 1690 input neurons, which are connected to the first dense hidden layer having 512 neurons followed by ReLU activation function. The second and third hidden layers contain 1024 and 512 neurons respectively both followed by the ReLU activation function. A dropout layer with a rate of 0.3 is then added to induce regularization and avoid overfitting. After the dropout layer, the values pass through two more hidden layers containing 128 and 64 neurons respectively along with the ReLU activation function again, and another dropout layer with a 0.2 rate. The final output layer having 20 neurons, equal to the total number of classes, terminates the neural network. Due to the problem being a multi-class classification, the output layer makes use of a Softmax function which formulates the final output probabilities of each class.
\subsection{Dataset Split}
The dataset was divided into training and validation or testing sets in the ratio 8:2 using stratified splitting, such that the number of examples from each of the 20 classes split proportionally into two sets. Stratifying was necessary due to the imbalanced classes, to avoid a disproportionate division of examples of classes with a relatively little or huge number of examples. The training and test sets contained 10,943 training examples and 2,736 test examples respectively after the split.
\subsection{Model Training}
The total number of trainable parameters that resulted from the proposed architecture is 1,991,124. The Adam optimizer was used with an initial learning rate of 0.0001 while training on the examples which were further divided into mini-batches of size 32, to implement mini-batch gradient descent with respect to the sparse categorical cross-entropy loss function. The training took place over 100 epochs. 

\begin{figure*}[ht!]
	\centering
    \includegraphics[width=\linewidth, height=4 cm]{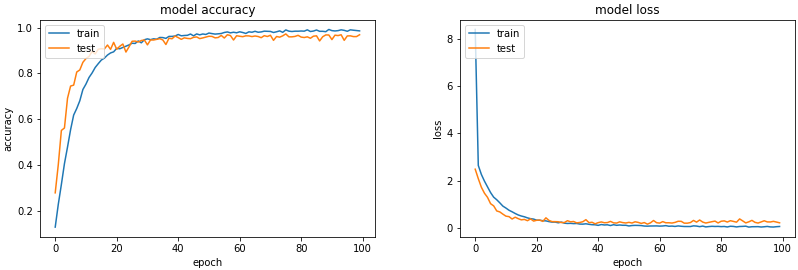}
    \caption{Accuracy and Loss on the Train and Test sets}
    \label{fig:acc_loss}
\end{figure*} 

\begin{figure*}[ht!]
	\centering
    \includegraphics[width= \textwidth,height=9cm]{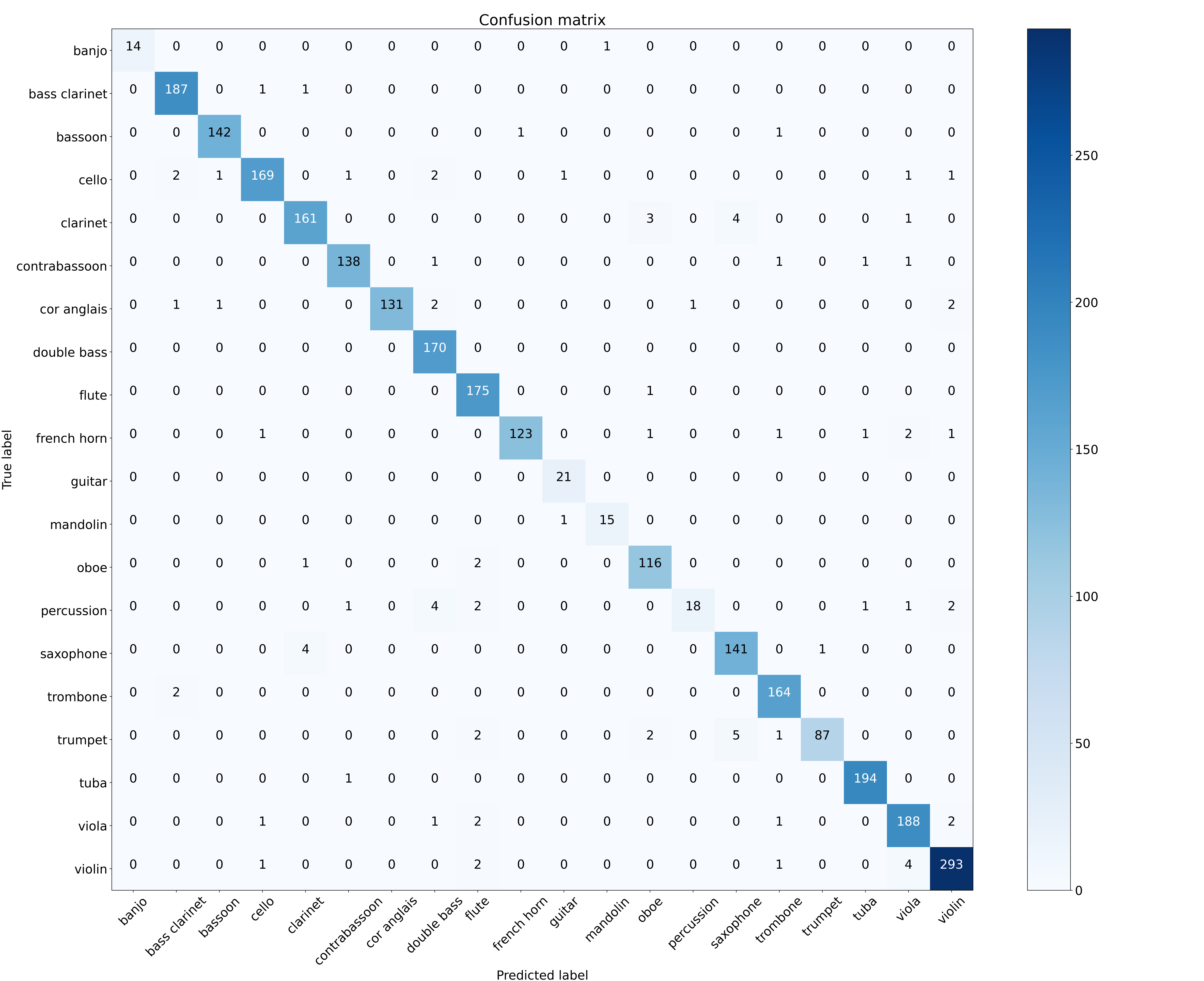}
    \caption{Confusion Matrix}
    \label{fig:cm}
\end{figure*}

\begin{figure*}[ht!]
	\centering
    \includegraphics[width=\textwidth, height=7cm]{ROCandPrecRec}
    \caption{Precision-Recall and AUC-ROC curves}
    \label{fig:PR}
\end{figure*}

\section{Results and Analysis}
\label{sec:results}

During model training, the training accuracy peaked 0.9913 and validation accuracy 0.9726.

Albeit a high and stable validation accuracy was obtained as shown in Figure \ref{fig:acc_loss}, it could not be concluded as the best metrics for the model evaluation since the dataset had highly imbalanced classes, as described in Section \ref{sec:dataset}. This often brings in the accuracy paradox \cite{valverde2014100}, i.e misclassifying minority class examples yet achieving a high accuracy due to correct classification of a relatively bigger number of majority class examples. 
Therefore, a confusion matrix was plotted and the F1 score of each class prediction were calculated. This would reveal the true evaluation of our model.

From the Confusion matrix shown in Figure \ref{fig:cm}, and F1 scores displayed in Table \ref{table:pr}, it is evident that the test minority class examples have been predicted with minimal error as well. Class `percussion' has a relatively lower F1 score, which is quite expected because, unlike the other classes, it consists of a very small number of examples from 39 different percussion instruments, as explained in Section \ref{sec:dataset}. 

\begin{table*}[ht!]
\centering
\caption{Precision, Recall and F1 Scores of each class}
\label{table:pr}
\begin{tabular}{l|llll}
\hline
              & \textbf{Precision} & \textbf{Recall} & \textbf{F1-Score} & \textbf{Support} \\
\hline
\hline
banjo         & 1         & 0.93   & 0.97     & 15      \\
bass clarinet & 0.99      & 0.98   & 0.98     & 189     \\
bassoon       & 0.97      & 1      & 0.99     & 144     \\
cello         & 0.99      & 0.94   & 0.97     & 178     \\
clarinet      & 0.98      & 0.95   & 0.96     & 169     \\
contrabassoon & 0.99      & 0.96   & 0.98     & 142     \\
cor anglais   & 0.99      & 0.99   & 0.99     & 138     \\
double bass   & 0.96      & 0.99   & 0.97     & 170     \\
flute         & 0.95      & 0.99   & 0.97     & 176     \\
french horn   & 0.98      & 0.97   & 0.98     & 130     \\
guitar        & 1         & 1      & 1        & 21      \\
mandolin      & 0.89      & 1      & 0.94     & 16      \\
oboe          & 0.96      & 0.96   & 0.96     & 119     \\
percussion    & 0.86      & 0.66   & 0.75     & 29      \\
saxophone     & 0.88      & 0.97   & 0.92     & 146     \\
trombone      & 0.99      & 0.98   & 0.99     & 166     \\
trumpet       & 0.96      & 0.92   & 0.94     & 97      \\
tuba          & 0.99      & 0.99   & 0.99     & 195     \\
viola         & 0.96      & 0.93   & 0.95     & 195     \\
violin        & 0.97      & 0.99   & 0.98     & 301     \\
\hline
accuracy      &           &        & 0.97     & 2736    \\
\hline
macro avg     & 0.96      & 0.96   & 0.96     & 2736    \\
weighted avg  & 0.97      & 0.97   & 0.97     & 2736   \\
\hline
\end{tabular}
\end{table*}

To further support the high accuracy of 97\%, the AUC-ROC and Precision-Recall curves were also plotted, as shown in Figure \ref{fig:PR}. An AUC score of 0.996 was achieved, which further supports the results, and permits to make the claim of the commendable results without being influenced by the accuracy paradox, which is commonly the case of high accuracy on imbalanced class problems. It can also be observed that `percussion' has comparatively worse Precision-Recall and AUC-ROC curves.

\section{Conclusion and Future Work}
\label{sec:conclusion}

In this paper, we formulated a baseline model to work on musical instrument recognition. The model surprisingly achieved a new state-of-the-art accuracy of 97\% on the full dataset containing all 20 classes of different musical instruments, despite the heavy class imbalance and the fact that most instruments belonged to a particular family viz. Strings, Woodwinds, Brass and Percussion. The experimental setup, Section \ref{sec:setup}, was finalized after a commendable number of iterations of hyperparameter tuning. Although a different set of number of layers and neurons had resulted in a slightly better validation accuracy, this particular model resulted in a better and more uniform F1 score over the classes in addition to a more stable fluctuation of the accuracy curves, as shown in Figure \ref{fig:acc_loss}. Additionally, the h5 file of the model consumes only 22.8 MB of memory. Thus, the proposed model can be incorporated with web and mobile applications with cheaper memory requirements.

Nevertheless, there are tremendous scopes for future work that may result in an even better performance on this particular dataset. Data augmentation measures may be adopted to deal with the imbalance problem. Different activation functions and optimizers with varying learning rates may be tried. MFCCs and mel-sprectograms provide excellent visual perceptions of sound, thus CNNs may prove to be quite efficient. There are a lot of variables that can be tweaked during the pre-processing stages as well, such as choosing a longer duration of the examples, a different frame size, hop length, or more number of MFCC coefficients. Expanding our target space by supporting the recognition of even more instruments including the piano, or the ukulele, for instance, would be a notable improvement on our current model.

\section*{Acknowledgements} 
We would like to thank the Department of Computer Science and Engineering and Center for Natural Language Processing (CNLP) at National Institute of Technology Silchar for providing the requisite support and infrastructure to execute this work. 

\FloatBarrier

\small{
\bibliographystyle{cys}
\bibliography{main}
}
\normalsize

\begin{biography}[]{} 
\end{biography}


\end{document}